\documentclass[aip,apl,amsmath,amssymb,reprint]{revtex4-1}

\usepackage{pgfplots}
\pgfplotsset{compat=newest}

\usepackage{dcolumn}% Align table columns on decimal point
\usepackage{bm}% bold math

\usepackage[utf8]{inputenc}
\usepackage[T1]{fontenc}
\usepackage{mathptmx}
\usepackage{etoolbox}
\usepackage{hyperref}

\makeatletter
\def\@email#1#2{%
 \endgroup
 \patchcmd{\titleblock@produce}
  {\frontmatter@RRAPformat}
  {\frontmatter@RRAPformat{\produce@RRAP{*#1\href{mailto:#2}{#2}}}\frontmatter@RRAPformat}
  {}{}
}%
\makeatother
\begin{document}

\title{Tunable integrated ring resonators by femtosecond laser micromachining} %Title of paper

\newcommand{\POLIMI}{Dipartimento di Fisica - Politecnico di Milano, piazza Leonardo da Vinci 32, 20133 Milano (Italy)}
\newcommand{\IFN}{Istituto di Fotonica e Nanotecnologie - Consiglio Nazionale delle Ricerche (IFN-CNR), piazza Leonardo da Vinci 32, 20133 Milano (Italy)}
\newcommand{\UNIPV}{Dipartimento di Fisica ``A. Volta'' - Universit\`{a} di Pavia, Via Bassi 6,
27100 Pavia (Italy)}
\newcommand{\EPHOS}{Ephos, Viale Decumano 34, 20157 Milano (Italy)}

\author{Giulio Gualandi}
\affiliation{\POLIMI}\affiliation{\IFN}

\author{Fabio Saretto}
\affiliation{\POLIMI}

\author{Daniele Pedroli}
\affiliation{\UNIPV}

\author{Giacomo Corrielli}
\affiliation{\IFN}\affiliation{\EPHOS}

\author{Marco~Liscidini}
\affiliation{\UNIPV}\affiliation{\IFN}

\author{Roberto~Osellame}
\affiliation{\IFN}\affiliation{\EPHOS}

\author{Andrea~Crespi}
\email{andrea.crespi@polimi.it}
\affiliation{\POLIMI}\affiliation{\IFN}

%\date{}

\begin{abstract}
Femtosecond Laser Micromachining (FLM) is a powerful technology for the fabrication of photonic devices. In this context, the integration of resonant elements within the platform represents a key advancement, enhancing both its versatility and its compatibility with a wide range of optical and fluidic components specifically enabled by this technique. Here, we report the realization of a tunable racetrack resonator fabricated by FLM and operating at telecom wavelengths. 
Leveraging low-loss waveguides, we obtained a Q factor of the resonator as high as $8 \times 10^5$ at critical coupling. Moreover, by integrating two thermo-optic phase shifters, we achieved both resonance tuning and dynamic control of the Q factor. This capability makes the device highly versatile for applications requiring dynamic spectral control, such as tunable filters, gyroscopes, and sensors. 
\end{abstract}

\maketitle 

\section{Introduction}

Optical resonators are fundamental building blocks in photonic integrated circuits (PIC), as their intrinsic properties make them suitable for a broad spectrum of applications, including filters \cite{geuzebroek2006ring, xu2021ultranarrow}, modulators \cite{yuan2024microring, hong2017triple}, sensors \cite{sarkaleh2017ring}, gyroscopes \cite{lai2020earth, khial2018nanophotonic}, and nonlinear optics enhancement \cite{tritschler2025nonlinear, tritschler2025squeezed}. When dealing with integrated dielectric systems, ring resonators are arguably among the most versatile and successful implementations. A ring resonator is essentially a waveguide bent into a closed loop that confines light to circulate within it. In this configuration, the resonant frequencies are determined by the interplay between the dispersion relation of the waveguide modes and the loop length. Coupling light into and out of the resonator is typically achieved through evanescent coupling, by placing one or more access waveguides in close proximity to the ring \cite{Rabus2020}.

The realization of ring resonators in a given photonic platform strongly depends on refractive index contrast and the adopted fabrication technology. For example, in silicon-on-insulator (SOI) photonic platforms, the large refractive index contrast allows one to fabricate very compact structures with radii of only a few micrometers at telecom wavelengths while still maintaining long photon dwelling times \cite{Bogaerts2012,Guo2024}. However, if the refractive index contrast decreases, bending losses increasingly limit the achievable light confinement, making it necessary to adopt larger radii. This is the case, for instance, when the material of choice is silicon nitride. In fact, with this material low-loss microrings typically have radii in the tens to hundreds of micrometers \cite{Lipson2017,Cui2023}.

\begin{figure}[!ht]
    \centering

    \begin{tikzpicture}[scale=1.1]
        \node at (0,0) {\includegraphics[width=8.25cm]{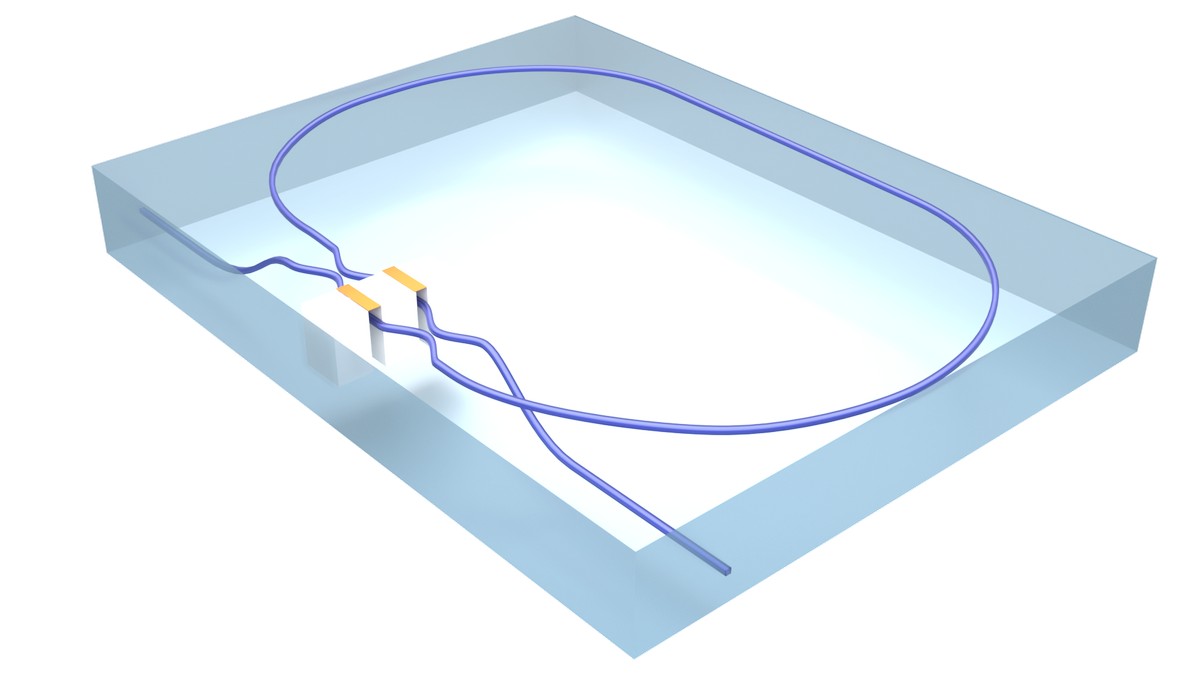}};
       \node at (-1.5,0.22) [below left] {$P_1$};
        \node at (-1.2,0.37) [above right] {$P_2$};

		\draw [blue,dashed] (-2.42,0.63)--++(-35:1.8)  node [below] {MZI}--++(23:0.8) --++(146:1.75)--cycle;

        \draw [very thick, red,->] (0.84,-1.48) ++(-34:0.1)--++(-34:0.5) node [right] {OUT};
        \draw [very thick, red,<-] (-2.89,0.82) ++(143:0.1)--++(143:0.5) node [above] {IN};
		 \draw [->] (0.83,0.26)--++(8:1.6) node [midway, below] {$R$};
    \end{tikzpicture}
 
    \caption{Layout of the resonator. The waveguide, starting from the input port (IN) first forms a ring with a racetrack geometry, then couples back to itself through a MZI, and is finally brought to the output (OUT) port, passing underneath the ring structure. The MZI is equipped with two TOPS, one per each arm, consisting in metal resistors deposited and patterned on the substrate surface. Different electrical powers $P_1$ and $P_2$ can be dissipated on the resistors, enabling independent control on the MZI phases. To enhance thermal efficiency, U-trenches were ablated around the waveguide segments \cite{Ceccarelli2020}. Total footprint of the device is $5 \times 6~\mathrm{cm}^2$.}
    \label{fig:layout}
\end{figure}
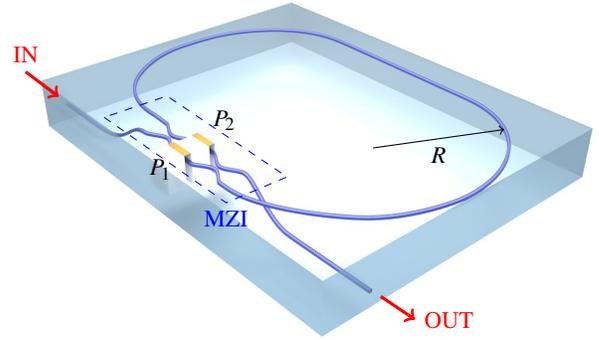

Femtosecond Laser Micromachining (FLM) has emerged over the last two decades as a versatile and powerful platform for the fabrication of photonic integrated circuits (PICs), offering intrinsic three-dimensional capabilities and rapid prototyping in glass substrates \cite{osellame2012,Gross2015,Sun2022}. In this material platform, however, the realization of ring resonators presents particular challenges, as FLM waveguides are formed by local refractive index modification induced by laser irradiation, for which the resulting index contrast is extremely low, typically in the range of $10^{-3}  - 10^{-4}$. This imposes the use of bending radii on the order of millimeters to centimeters to mitigate radiation losses. Moreover, when inscribing a closed loop, undesired inhomogeneities may arise at the overlap region where the beginning and end of the irradiated track meet, further complicating the fabrication of high-quality resonators. Such technological limitations have so far hindered the widespread demonstration of FLM-based ring resonators.

In this work, we report the realization and experimental characterization of reconfigurable integrated optical resonators fabricated by FLM providing a design approach to overcome the issues associated with inscription inhomogeneities. Our devices are realized according to the layout depicted in Fig.~\ref{fig:layout}, with a single waveguide that couples to itself through a Mach-Zehnder interferometer (MZI), and then reaches the output by passing below the ring structure. Two thermo-optic phase shifters (TOPSs) are realized on the internal arms of the MZI. By controlling the thermal powers $P_1$ and $P_2$ dissipated on the TOPS it is possible to actively tune both the wavelength shift of the spectral resonances and the Q factor of the device.

\begin{figure*}[t]
    \centering
    \includegraphics[width=1\textwidth]{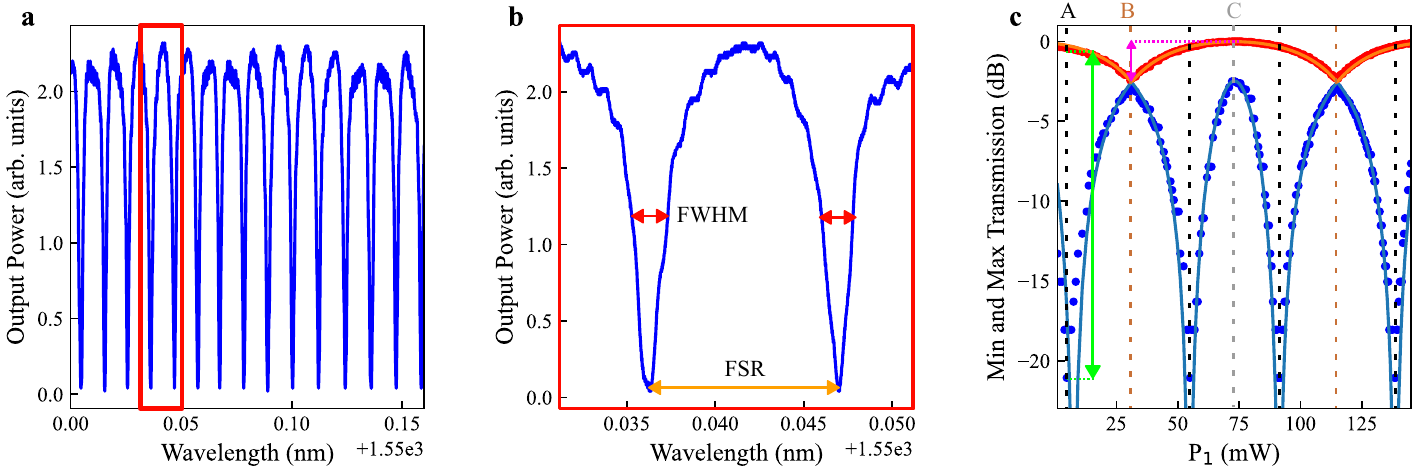}
    \caption{
    Characterization of the spectral response of the resonator. 
    (a) Output power of the resonator as a function of wavelength under critical coupling conditions. Multiple resonance dips are visible over a span of 160\,pm.
    (b) Zoom-in of the spectral region highlighted by the red rectangle in (a), showing the shape and depth of individual resonance fringes, with the FWHM and FSR indicated by red and orange arrows, respectively.
    (c) Maximum (red dot) and minimum (blue dot) transmission values extracted from the spectra as $P_1$ is varied, thereby tuning the coupling coefficient \( t \). Solid lines, represent a theoretical fit, taking into account unbalanced directional couplers, see text. The dotted line labeled A (black), B (brown), C (grey) correspond to $|t|=\alpha$ (critical coupling condition), strong undercoupling ($|t| \sim 1$) and strong overcoupling ($|t| \sim 0$), respectively. The purple arrow indicates the internal losses of the ring, while the green arrow marks the extinction ratio achieved experimentally at critical coupling. 
    }
    \label{fig:Figure2_Fringes}
\end{figure*}

\section{Device design and fabrication}

The device design exploits the three-dimensional capabilities of FLM to realize a monolithic resonator formed from a single physically open waveguide that simultaneously functions as the resonator and as the input–output channel. In this way, one avoids the additional losses that can occur at the junction when fabricating fully closed loops with FLM. Furthermore, by exploiting the three-dimensionality of the waveguide paths, we also demonstrate the fabrication of several concentric resonators within the same glass chip: the straight sections of the waveguides are inscribed at a depth of 55~$\mu$m below the glass surface, while the bent sections forming the rings are inscribed at 35~$\mu$m, enabling vertical stacking without optical crosstalk. Additional details on the device geometry are provided in Appendix~\ref{secApp:Design}.

We fabricated 20 concentric resonators in a single substrate of Corning Eagle XG glass of $5 \times 6~\mathrm{cm}^2$ footprint, using curvature radii $R$ in the range between 16.8~mm and 21.0~mm. The waveguides were inscribed by four overlapped laser scans, translating the substrate with respect to the laser focus using 3-axis air-bearing translators (Aerotech ABL1500) with sub-micron accuracy. In detail, we used laser pulses of 480-nJ energy, 180~fs duration and 1~MHz repetition rate from a commercial system (LightConversion Pharos), focused by a 0.5~NA water-immersion objective. Two different translations speeds were employed (5~mm s$^{-1}$ and 7~mm s$^{-1}$). After laser irradiation, the substrate underwent thermal annealing post-processing to improve the optical properties of the waveguides \cite{Arriola2013, Corrielli2018}. These fabrication parameters ensure single-mode guiding at 1550~nm wavelength with a mode diameter ($1/e^2$) of about 7.5~$\mu$m, and propagation loss <0.2 dB/cm. TOPS were fabricated as in Ref.~\onlinecite{Ceccarelli2020}: U-shaped trenches were defined around the MZI arms, using water-assisted laser ablation, to provide thermal insulation; a 100-nm Cr/Au layer was deposited onto the substrate and the resistor geometry was patterned by femtosecond laser ablation. Such microstructured layout enable high-efficiency actuation (with typical electrical power values on the order of tens of mW), low cross-talk between the two TOPS, and reduces the heated area on the substrate, thus limiting thermal drifts of the optical properties of the resonator. The device was finally glued to a custom aluminum basis to improve thermal stability.

\section{Device operation theory}

We now describe in more detail the operating principles of the device sketched in Fig. \ref{fig:layout}. At its core lies an MZI that functions as a reconfigurable coupler, operating on the two distinct input waveguide modes a transformation described by the matrix $U=[u_{ij}]$. In particular, the coupling properties are governed by the phase shifts $\phi_1$ and $\phi_2$ accumulated in the two arms, which can be controlled by varying the corresponding TOPSs.  In the ideal case of perfectly balanced (i.e. 50:50) directional couplers and negligible losses, the cross-coupling coefficient of the MZI is equal to:
\begin{equation}
u_{12} = u_{21} = t e^{j\phi_\text{cm}} = \cos\left(\frac{\Delta\phi}{2} \right) e^{j\phi_\text{cm}},  \label{eq:MZI} 
\end{equation}
where $\phi_{\text{cm}} =\frac{1}{2}( \phi_1 + \phi_2+\pi)$ is the total phase accumulated in the MZI, and $\Delta\phi = \phi_1 - \phi_2$ is the phase difference between the two MZI arms (see Appendix~\ref{secApp:TransfFunction}). By taking into account the thermal crosstalk between the two TOPSs, the dependence of  $\phi_{\text{cm}}$ and $\Delta\phi$ on the dissipated electrical power $P_1$ and $P_2$ can be expressed as~\cite{flamini2015thermally}:
\begin{align}
\Delta\phi &= \frac{2\pi}{\lambda_0}\left[a_1 P_1 + a_2 P_2 + \Delta L\right], \label{eq:delta_phase}\\
\phi_\text{cm} &= \frac{2\pi}{\lambda_0}\left[b_1 P_1 + b_2 P_2 + L\right]+\frac \pi 2,
\label{eq:common_phase}
\end{align}
where $\lambda_0$ is the light wavelength in vacuum, $a_{i}$ and $b_{i}$ are coefficients depending on the device geometry and material properties, $L$ is an effective optical path length of the MZI (thus taking into account also the effect of the refractive index) and $\Delta L$ is the  optical-length imbalance of the MZI arms introduced by fabrication tolerances.

The power transmission of the device is given by (see \ref{secApp:TransfFunction} and Ref.~\onlinecite{yariv2000universal}):
\begin{equation}
\mathcal{T} = \frac{\alpha^2 + t^2 - 2\alpha t \cos(\theta_R+\phi_\text{cm})}{1 + \alpha^2 t^2 - 2\alpha t \cos(\theta_R+\phi_\text{cm})},
\label{eq:T}
\end{equation}
where $\alpha$ is the amplitude transmission per round trip and $\theta_R$ is the phase accumulated per round-trip excluding the MZI section. Finally, it is convenient to associate the round-trip phase appearing in \eqref{eq:T} with an effective total optical path length $L_\mathrm{tot}$ defined as
\begin{equation}
\theta_R +\phi_{cm} = \frac{2 \pi}{\lambda_0} L_\mathrm{tot}.
\end{equation}
Since in our structure the round trip in the resonators includes the MZI, $L_\mathrm{tot}$ depends also on $P_1$ and $P_2$ (see \eqref{eq:common_phase}) and is not simply linked to the physical length of the device. Yet, as in the case of conventional ring resonators, a resonant wavelength $\lambda_{0,m}$ satisfies the relation $m\lambda_{0,m}=L_\mathrm{tot}$, with $m$ indicating the resonance order. Similarly, the device transmission shows dips that are, for a sufficiently small spectral range, periodic in wavelength with a free spectral range $\text{FSR} = \frac{\bar{\lambda}_0^2}{L_\mathrm{tot}}$,
where $\bar{\lambda}_0$ is the mean wavelength in vacuum in the interval of interest (here, it can safely be taken as $\bar{\lambda}_0$ = 1550~nm).

\begin{figure*}
  \centering
  \includegraphics[width=1\textwidth]{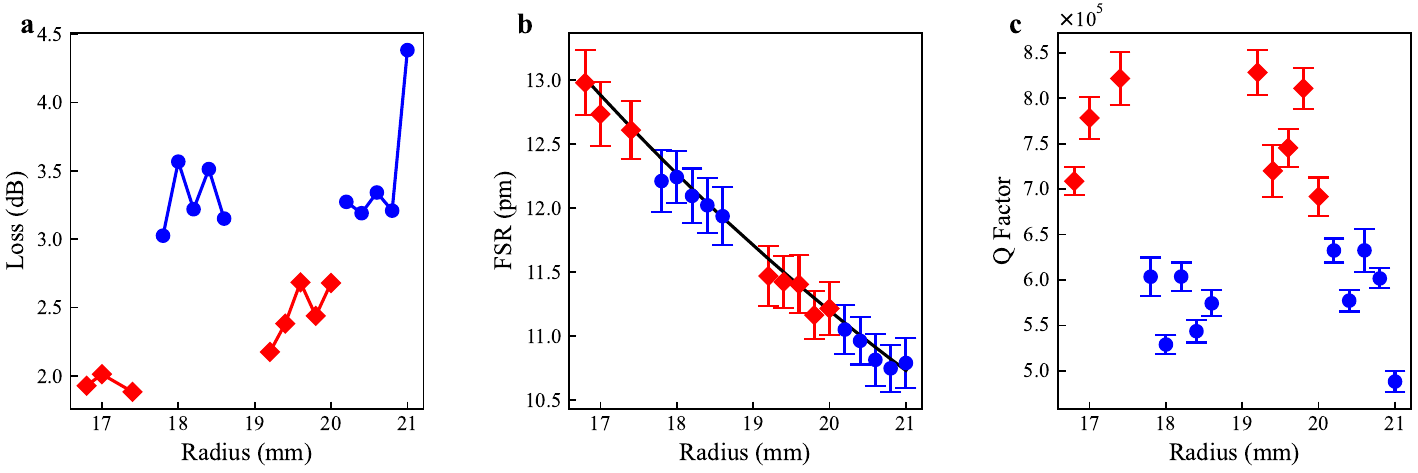}
  \caption{
    Characterization of all resonators as a function of the ring radius, where blue dotted markers represent the values obtained with a writing speed of 5~mm s$^{-1}$, while red diamonds correspond to those fabricated at 7~mm s$^{-1}$.
    (a) Measured round-trip losses of the ring resonator. 
    (b) Free Spectral Range (FSR), where the green line represents the theoretical prediction based on the relation $\text{FSR} = \frac{\bar{\lambda}^2}{L_\mathrm{tot}}$, and the data points indicate measured values, calculated by averaging the separation between resonance dips detected over the scanned wavelength range.
    (c) Quality factor of each resonator, extracted from the FWHM under critical coupling conditions. 
  }
  \label{Figure3_Characterization}
\end{figure*}

Another figure of merit characterizing the performance of optical resonators is the $Q$ factor, defined as the ratio between the resonance frequency and the full width at half maximum (FWHM) of the corresponding spectral dip. In our case one can show that~\cite{Yang2023}:
\begin{equation}
Q = \frac{\lambda_0}{\text{FWHM}} = \frac{\pi L_\mathrm{tot}}{\lambda_0} \left(\arccos\frac{2\alpha |t|}{1 + \alpha^2 |t|^2}\right)^{-1} \label{eq:Q}.
\end{equation}
This parameter is intrinsically related both to the round-trip losses and to the optical length of the loop. 

It is worth noting that, while the common phase $\phi_{\text{cm}}$  determines a shift in the resonances, the phase difference $\Delta \phi$ governs $|t|$ and thus modulates the depth of the transmission dips and the $Q$ factor. Critical coupling is achieved when the MZI transmission equals the ring transmission, namely $|t|=\alpha$; in this condition complete destructive interference ($\mathcal T =0$) is achieved at resonance.

\section{Experimental results}

All the fabricated devices were characterized by measuring the transmission spectrum in different coupling conditions, namely with different values of $|t|$. In detail, horizontally-polarized coherent light from a tunable C-band laser source (Santec TLS-570) was coupled to the device input using an aspherical lens (0.18 NA). The transmitted light is collected from the output port using another aspherical lens (0.4 NA) and directed to an amplified photodiode (Thorlabs PDA10CS2). The laser wavelength was continuously swept over a 160~pm range starting from 1550.000 nm at 1~nm/s speed, while the photodiode signal was recorded by a digital oscilloscope. An example of the resulting transmission spectrum is shown in Fig.~\ref{fig:Figure2_Fringes}a. To study different coupling conditions, corresponding to different values of $|t|$, we measured the transmission for different values of the dissipated power $P_1$ on the first TOPS, leaving $P_2=0$.

From the acquired spectrum we can experimentally retrieve both the FWHM and the FSR of the transmission dips (see also Fig.~\ref{fig:Figure2_Fringes}b), as averages of the widths and of the distances of the resonance dips found in the acquired wavelength range.

In addition, from each spectrum we extracted the minimum ($V_{min}$) and maximum values ($V_{max}$) of the signal.  Fig.~\ref{fig:Figure2_Fringes}c plots $V_{min}$ and $V_{max}$ as a function of $P_1$ (respectively, the blue and red curves), as an example in the case of the device with $R$~=~19.8~mm. The maximum and minimum of the red curve correspond to the cases of minimal $|t|\sim1$ and maximal $|t|\sim 0$ coupling to the ring respectively. Importantly, these values can be used to estimate experimentally the internal round-trip losses of the resonator using:
\begin{equation}
\left.\frac{1}{\alpha^2}\right|_{\text{dB}} = -10 \log_{10} \left( \frac{\min\lbrace V_{\text{max}}\rbrace}{\max \lbrace V_{\text{max}} \rbrace} \right).
\label{eq:alpha}
\end{equation}

The measured losses are reported in Fig.~\ref{Figure3_Characterization}a for all fabricated resonators. The increase in loss with the curvature radius is primarily due to the longer loop length, showing that propagation losses in the waveguides are dominant with respect to additional losses due to bending, for the tested radii. Waveguides written at a speed of 7~mm s$^{-1}$ demonstrate significantly better performance, consistently remaining below 3~dB overall loss. In particular, the device with $R$~=~17.4~mm yields the minimum loss of about 1.9~dB, over a loop geometrical length of 127.6~mm. 

The vertical distance between the blue and the red curve in the graph of Fig.~\ref{fig:Figure2_Fringes}c gives the extinction ratio of the transmission dips for a given value of $P_1$. The maximum extinction ratio is achieved in correspondence of the critical coupling points, where $|t|=\alpha$. If the directional couplers of the MZI were perfectly balanced, by varying $P_1$ we would be able to scan the full interval of values of $|t|$ from 0 to 1 following precisely ~\eqref{eq:MZI}; in this ideal case, the maxima of the blue curve would touch the red curve, resulting from a flat transmission spectrum at those points. The distance between the curves still observed in those points indicates that $|t|=0$ and $|t|=1$ cannot be achieved precisely due to imperfections in the directional couplers of the MZI.

In fact, we also report as a solid line the fit from a theoretical model which includes the possibility of unbalanced couplers (but still equal one to the other) composing the MZI. As shown in Fig.~\ref{fig:Figure2_Fringes}c, the experimental points are well reproduced by assuming 64:36 power splitting ratio of the individual couplers (see Appendix~\ref{secApp:Fit} for additional details). This improved theoretical model would still predict vanishing transmission at the minima, whereas in practice a finite residual background is observed. This is likely due to the limited spectral bandwidth of the tunable laser; other small deviations between the data and the fitted curve are attributable to the non uniform wavelength scanning of the laser.

Figure \ref{Figure3_Characterization}b reports the measured FSR values for all devices, compared to a theoretical curve calculated as $FSR = \bar{\lambda}_0^2/{L_\mathrm{tot}}={\bar\lambda}_0^2/(n \mathcal L)$, with $n$~=~1.49 \cite{Cushman2016EagleXG} and with $\mathcal L$ the nominal geometrical length of the waveguide loop (counting both the length of the race-track and the length of the MZI waveguides).

Finally, Fig.~\ref{Figure3_Characterization}c reports the quality factor $Q_\mathrm{cc}$ at the critical coupling condition for all the fabricated devices. The quality factor at the critical coupling is $Q_\mathrm{cc}=Q_{i}/2$, with $Q_{i}$ the intrinsic quality factor, and hence it depends only on the propagation losses per unit of length $\alpha_L$ of the ring waveguide according to $Q_\mathrm{cc}=\pi n_g/(\lambda_0\alpha_L)$ (see Appendix~\ref{secApp:Qfactor} for additional details). Thus, from the results shown in Fig.~\ref{Figure3_Characterization}c, one can directly estimate the propagation losses per unit of length, which are about $0.23\pm0.02$ dB/cm for $5$ mm/s writing speed and of about $0.17\pm0.02$ dB/cm for 7 mm/s.  Such losses are essentially independent of radii, with small fluctuation from sample to sample. This indicates negligible bending losses for radii larger than 17 mm for both fabrication speed. As a further check, inverting \eqref{eq:Q} with these $Q$ factor values yields round-trip losses in good agreement with those previously determined.
The best-performing resonator exhibits $Q_\mathrm{cc}\approx 8.3 \times 10^5$ corresponding to $Q_i \approx 1.6 \times 10^6$. 

\begin{figure}
    \centering
    \includegraphics[width=8cm]{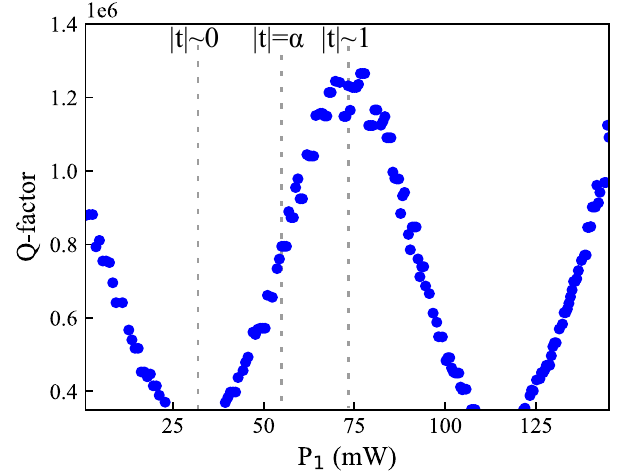}\hspace{1em}
    
    \caption{Q factor as a function of dissipated power $P_1$, illustrating the transition between coupling regimes, from strong overcoupling ($|t| \sim 0$) to strong undercoupling ($|t| \sim 1$), passing from critical coupling ($|t|=\alpha$).
    The highest Q factor is observed in undercoupling regime. No data points are shown at the minima, as the transmission spectrum becomes flat in those regions, preventing the identification of a resonance dip and, consequently, the measurement of a meaningful FWHM.}
    \label{fig:Figure4_Qfactor}
\end{figure}

As mentioned, the $Q$ factor changes with the coupling condition defined by the MZI, that in our device can be tuned actively. Fig.~\ref{fig:Figure4_Qfactor} shows the $Q$ factor as a function of the dissipated power $P_1$, calculated from the FWHM of the spectral dips, in the case of the device with $R$~=~19.8~mm. The plot highlights three notable coupling conditions: critical coupling ($|t| = \alpha$), strong overcoupling ($|t| \approx 0$) and strong undercoupling ($|t| \approx 1$). The highest $Q$ factor is observed in the undercoupling regime.

\begin{figure}

   \centering
    
    \includegraphics[width=8cm]{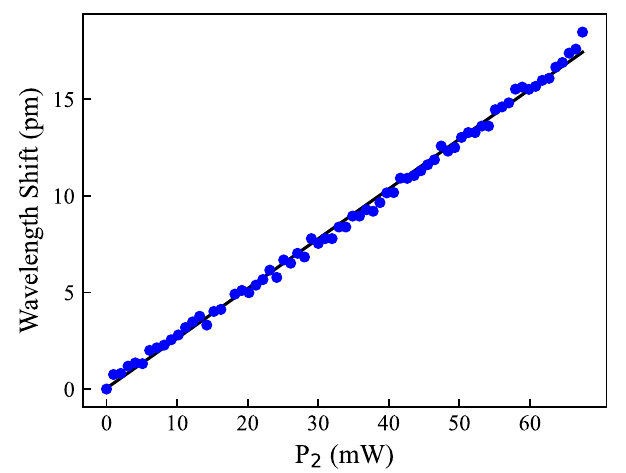}\hspace{1em}
    
    \caption{Diagram showing the resonance shift as a function of the power dissipated $P_2$. The data show a clear linear trend, highlighted by the black linear fit.}
    \label{Figure5_Convolution}
\end{figure}

Finally, we show that by simultaneously actuating both TOPSs on the two MZI arms we can tune the position of the resonances. We scan the power $P_2$ on the second TOPS, while adjusting at the same time $P_1$ according to:
\begin{equation}
P_1 = P_{1,\text{cc}} - \frac{a_2}{a_1} P_2 ,
\label{eq:P}
\end{equation}
where $P_{1,\text{cc}}$ is the dissipated power that produces the critical coupling condition when $P_2 = 0$ (namely, $\frac{2 \pi}{\lambda_0}\left(a_1P_{1,\text{cc}} + \Delta L\right)=\Delta \phi_\text{cc}$ that gives $|t|=\alpha$). In this way, we ensure that the phase shift difference $\Delta\phi$ in the MZI remains constant, thereby maintaining the critical coupling condition. At the same time, the common phase $\phi_{\text{cm}}$ is varied, effectively modifying the total phase accumulated in the ring. 
The values of the coefficients $a_1$ and $a_2$ were estimated by a preliminary characterization of the response of each TOPS separately (See Appendix~\ref{secApp:estCoeff} for further detail). For each value of $P_2$, we recorded the transmission spectrum, which was then convoluted with a reference one acquired with $P_2=0$, to accurately extract the wavelength shift of the dips. Results are showsn in Fig.~\ref{Figure5_Convolution}, together with a linear fit. We are able to change the position of the peak by more than 15~pm, namely more than one FSR, in this case in the order of 11~pm. This demonstrates complete tunability of the device.

\section{Conclusions}

We have demonstrated the tunable integrated ring resonators fabricated using FLM, reaching a $Q$ factor above $8\times 10^5$ at critical coupling. 
By actuating judiciously the two TOPS on the MZI that couples the ring resonator to the waveguide, we demonstrate precise tuning of the resonance position over more than one FSR, and the capability to change the $Q$ factor actively.
This capability makes our device versatile and of potential interest for applications requiring dynamic control of the spectral response, including tunable optical filters, gyroscopes, and other sensing systems.
Due to the inherently low thermo-optic coefficient of silica glass, and to the high control on the heat diffusion in our microstructured TOPS, this platform may offer improved thermal stability compared to silicon-based technologies.
Notably, this demonstration paves the way to innovative devices, integrating ring resonators with other three-dimensional optical or fluidic devices uniquely enabled by the capabilities of the FLM technology.

\begin{acknowledgments}
G.C. and R.O. acknowledge funding from the European Innovation Council through the EIC Transition project FUTURE (g.a. 101136471).
\end{acknowledgments}

\section*{Authors Declarations}

\subsection*{Conflict of interest}

M.L. and R.O. have a patent application related to the device described in this article. All other authors have no conflicts to disclose.

\subsection*{Author Contributions}

\textbf{Giulio Gualandi}: Formal Analysis (equal); Investigation (lead); Writing – Original Draft Preparation (equal); Writing – Review \& Editing (equal). 
\textbf{Fabio Saretto}: Formal Analysis (equal); Investigation (lead); Writing – Review \& Editing (equal). 
\textbf{Daniele Pedroli}: Formal Analysis (equal); Writing – Review \& Editing (equal). 
\textbf{Giacomo Corrielli}: Conceptualization (supporting); Funding Acquisition (equal); Investigation (supporting); Writing – Review \& Editing (equal). 
\textbf{Marco Liscidini}: Conceptualization (lead); Formal Analysis (equal); Supervision (equal); Writing – Review \& Editing (equal). 
\textbf{Roberto Osellame}: Conceptualization (lead); Funding Acquisition (equal); Supervision (equal); Writing – Review \& Editing (equal). 
\textbf{Andrea Crespi}: Conceptualization (supporting); Formal Analysis (equal); Investigation (supporting); Writing – Original Draft Preparation (equal); Writing – Review \& Editing (equal).

\section*{Data availability}

Raw data at the basis of the graphs in the Main Text, together with the Python scripts to generate them, are publicly available at \url{https://doi.org/10.5281/zenodo.17657883}.

\section*{References}

\bibliography{FLMringResonator}

%%%%%%%%%% APPENDIX

\cleardoublepage

\onecolumngrid

\appendix

\renewcommand\thefigure{\thesection\arabic{figure}} 
\setcounter{figure}{0}  

\section*{Appendix}

\section{Details of race-track resonator layout}

\label{secApp:Design}

\begin{figure}[b!]
\centering
\includegraphics[width=0.7\linewidth]{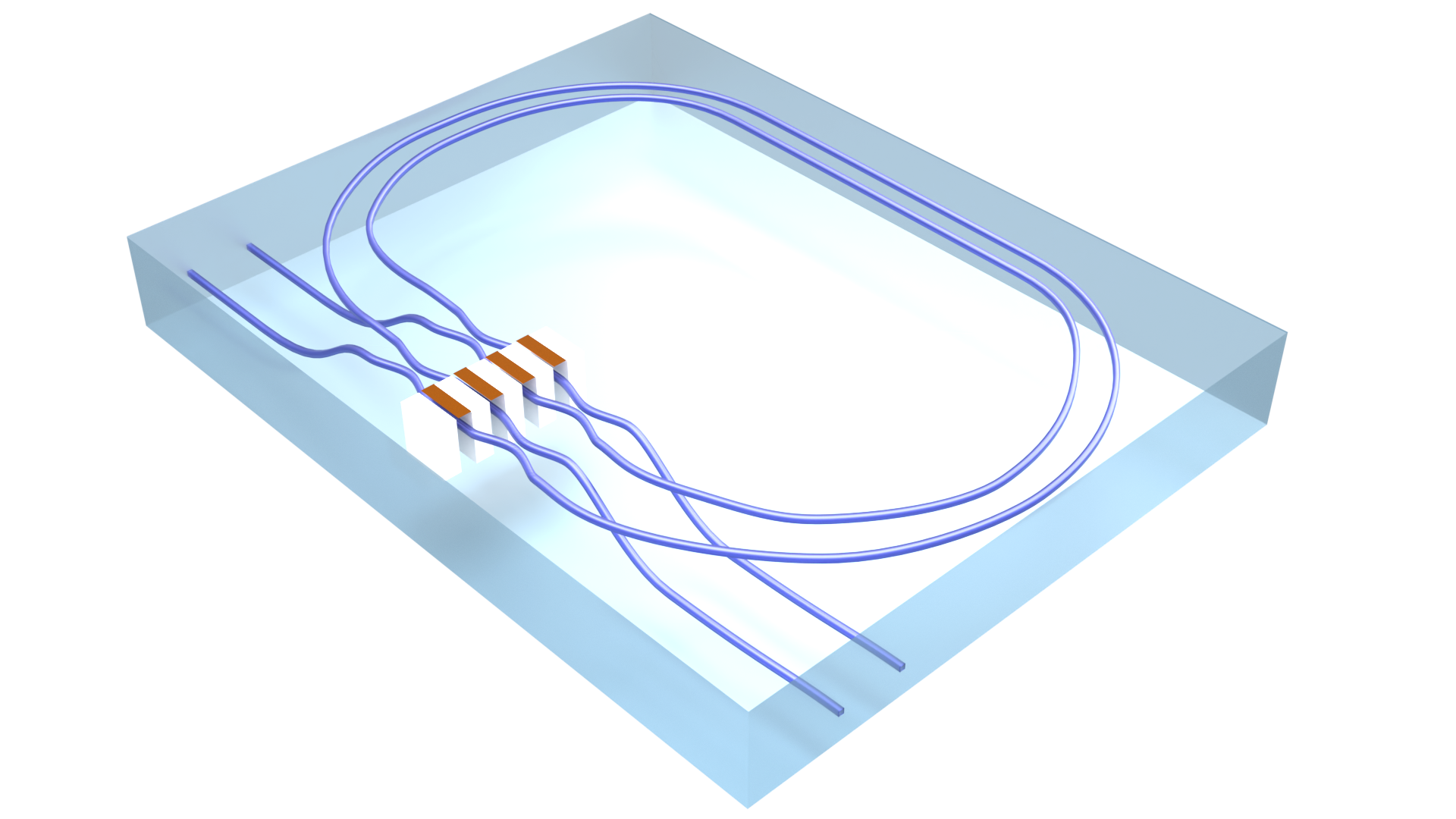}
\caption{Concentric resonator arrangement enabled by the three-dimensional capabilities of FLM. In total, 20 concentric rings were inscribed, with the racetracks and MZIs placed at a depth of 35~$\mu$m, while the input and output straight waveguides were written at 55~$\mu$m. For clarity, only two resonators are shown here. This design avoids waveguide crossings, as the paths pass at different depths, thereby preventing cross-talk and additional losses.}
\label{fig:resonator_geometry}
\end{figure}
The adopted geometry for the ring resonator,
depicted in Fig. 1 of the Main Text, consists of a single waveguide that couples to itself through a MZI. 

The input straight segment
is inscribed at a depth of 55~$\mu$m and propagates for 23~mm before gradually rising, through a spline curve with a minimum bending radius of 30~mm, to a final depth of 35~$\mu$m. This corresponds to a vertical displacement of 20~$\mu$m, implemented with a smooth transition that minimizes additional insertion losses.

The first side of the MZI is formed by two arc-sine curves with a minimum bending radius of 30~mm. Between these two curves, 
a 1-mm straight segment is included to accommodate the thermo-optic phase shifter with its microheater. The total length of the MZI, measured from the beginning of the first coupler to the end of the second, is 9.23~mm.

Following the MZI the waveguide continues into the racetrack section. This is drawn as a semicircumference of radius $R$, which is the one indicated as the characteristic radius of the resonator in our discussion e.g. in Fig. 1 and 3 of the Main Text. This first arc is  followed by another straight section equal in length to the MZI. The path then continues with a second circular arc, this time with a radius reduced by $40~\mu m$, corresponding to half of the displacement between the two arms of the MZI.  

Next, the waveguide continues by defining a second pair of arc-sine bends precisely aligned with the initial ones of the MZI, thereby guaranteeing the proper recombination of the optical paths. 

Finally, the waveguide crosses the race-track underneath, mirroring the same initial spline curve, thus returning to its original depth in the substrate. A final straight segment brings the waveguide to the output facet of the substrate.

This  design enables the integration of multiple resonators within a single piece of substrate by inscribing them concentrically, decreasing the radius at each iteration while keeping unchanged the length of the straight section of the racetrack (and therefore of the MZI). This concentric arrangement is illustrated in Fig.~\ref{fig:resonator_geometry}. In the figure only two  resonators are shown, for the sake of clarity; in fact, in this work we fabricated in the same substrate 20 concentric devices using this approach, with radii ranging from 21.0~mm down to 16.8~mm. 
Note that the waveguides of distinct devices never cross but rather pass over each other at different depths, thus avoiding cross-talk and additional losses.

\section{Transfer function of the device}
\label{secApp:TransfFunction}

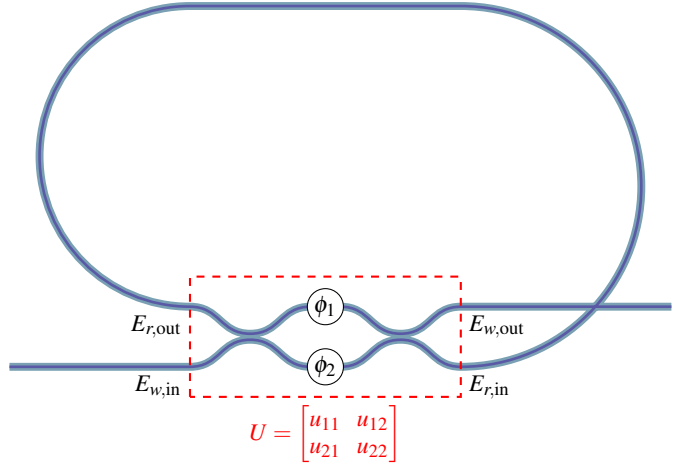
\begin{figure}
\centering
\begin{tikzpicture}[scale=0.8]
\draw [cyan!30!gray,line width=3pt] (-3,0)--(0,0) ..controls +(0.5,0) and +(-0.5,0) .. ++(1,0.45) ..controls +(0.5,0) and +(-0.5,0) .. ++(1,-0.45) --++(0.5,0) ..controls +(0.5,0) and +(-0.5,0) .. ++(1,0.45)  ..controls +(0.5,0) and +(-0.5,0) .. ++(1,-0.45) arc [start angle=-90, end angle=90, radius=3] --++(-4.5,0) arc [start angle=90, end angle=270, radius=2.5]  ..controls +(0.5,0) and +(-0.5,0) .. ++(1,-0.45) ..controls +(0.5,0) and +(-0.5,0) .. ++(1,0.45) --++(0.5,0) ..controls +(0.5,0) and +(-0.5,0) .. ++(1,-0.45)  ..controls +(0.5,0) and +(-0.5,0) .. ++(1,0.45)  --++(3.5,0);

\draw [blue!30!gray,line width= 1pt] (-3,0)--(0,0) coordinate (WIN) ..controls +(0.5,0) and +(-0.5,0) .. ++(1,0.45) ..controls +(0.5,0) and +(-0.5,0) .. ++(1,-0.45) --++(0.5,0) ..controls +(0.5,0) and +(-0.5,0) .. ++(1,0.45)  ..controls +(0.5,0) and +(-0.5,0) .. ++(1,-0.45) coordinate (RIN) arc [start angle=-90, end angle=90, radius=3] --++(-4.5,0) arc [start angle=90, end angle=270, radius=2.5] coordinate (ROUT) ..controls +(0.5,0) and +(-0.5,0) .. ++(1,-0.45) ..controls +(0.5,0) and +(-0.5,0) .. ++(1,0.45) --++(0.5,0) ..controls +(0.5,0) and +(-0.5,0) .. ++(1,-0.45)  ..controls +(0.5,0) and +(-0.5,0) .. ++(1,0.45) coordinate (WOUT) --++(3.5,0);

\node at (WIN) [below left] {$E_{w,\mathrm{in}}$};
\node at (WOUT) [below right] {$E_{w,\mathrm{out}}$};
\node at (RIN) [below right] {$E_{r,\mathrm{in}}$};
\node at (ROUT) [below left] {$E_{r,\mathrm{out}}$};

\draw [red,dashed,thick] (0,-0.5) rectangle (4.5,1.5);

\node at (2.25,-0.5) [below,red] {$U = \begin{bmatrix} u_{11} & u_{12}\\  u_{21} &  u_{22} \end{bmatrix}$};

\filldraw [draw=black, fill=white] (2.25,1) circle [radius=0.3] node {$\phi_1$};
\filldraw [draw=black, fill=white] (2.25,0) circle [radius=0.3] node {$\phi_2$};

\end{tikzpicture}

\caption{Concept scheme (not in scale) of our femtosecond-laser written racetrack resonator.  The MZI acts as a reconfigurable coupler implementing a transformation matrix $U$ on the modes, governed by the internal phases ($\phi_1$ and $\phi_2$ also indicated in the drawing). The field $E_{w,\mathrm{in}}$ enters the MZI and may be partially coupled to the ring when exit the same waveguide mode ($E_{r,\mathrm{in}}$). After propagating along the ring, the field $E_{r,\text{out}}$ is fed back into the other input mode of the MZI. The output mode of the MZI corresponding to the output port of the full device is $E_{w,\mathrm{out}}$.}
\label{fig:Fig_layout_description}
\end{figure}

\subsection{The coupled-resonator transfer function}

We report here the derivation of Eq. (4) of the Main Text, which relates the power transmission of the whole device to the parameters of the field-coupling operated by the Mach-Zehnder interferometer (MZI), and to the phase and loss figures of the resonator loop.

Coupling between the input and output fields of the MZI may be described by the matrix relation:
\begin{equation}
\begin{bmatrix}
E_{w,\mathrm{out}} \\
E_{r,\mathrm{in}}
\end{bmatrix}
= U  \begin{bmatrix}
E_{w,\mathrm{in}} \\
E_{r,\mathrm{out}}
\end{bmatrix} = 
\begin{bmatrix}
u_{11} & u_{12} \\
u_{21} & u_{22}
\end{bmatrix} \begin{bmatrix}
E_{w,\mathrm{in}} \\
E_{r,\mathrm{out}}
\end{bmatrix}
=
e^{j\phi}
\begin{bmatrix}
t & jk \\
jk^{*} & t
\end{bmatrix} \begin{bmatrix}
E_{w,\mathrm{in}} \\
E_{r,\mathrm{out}}
\end{bmatrix}
, \label{eq:matrRef}
\end{equation}
where $E_w$ and $E_r$ are the complex field amplitudes in the part of waveguide exiting directly the resonator and in the part of waveguide that travels the racetrack (see labels in Fig.~\ref{fig:Fig_layout_description}), while $t$ and $k$ denote the transmission and reflection coefficients and $\phi$ is a phase term. For a judicious choice of the physical boundaries of the MZI (namely including a little more or a little less of the input and output waveguide segments) and of the phase term $\phi$, we can always take $t$ as real while we will allow $k$ to be complex. The condition for $U$ to be a unitary matrix implies $t^2 + k k^* = 1$. 

In addition, the transmission of the loop is given by:
\begin{equation}
E_{r,\mathrm{out}} = \alpha e^{j \theta_R} E_{r,\mathrm{in}} \label{eq:loop}
\end{equation}
with $\alpha \in \mathbb{R}$ accounting for the loss of the loop and $\theta_R \in \mathbb{R}$ accounting for the acquired phase.

Substituting \eqref{eq:loop} in \eqref{eq:matrRef} we obtain:
\begin{gather}
E_{r,\mathrm{in}} = jk^* e^{j\phi} E_{w,\mathrm{in}}  + t e^{j\phi} E_{r,\mathrm{out}} = jk^*e^{j\phi} E_{w,\mathrm{in}} + t  \alpha e^{j (\theta_R+\phi)} E_{r,\mathrm{in}} \\
E_{r,\mathrm{in}} = \frac{jk^*e^{j\phi}}{1-t  \alpha e^{j (\theta_R+\phi)}} E_{w,\mathrm{in}}
\end{gather}
and again exploiting \eqref{eq:matrRef}:
\begin{multline}
E_{w,\mathrm{out}} = t e^{j\phi} E_{w,\mathrm{in}}+jk e^{j\phi} E_{r,\mathrm{out}} = t e^{j\phi} E_{w,\mathrm{in}}+ jk \alpha e^{j (\theta_R+\phi)} E_{r,\mathrm{in}} =\\= t e^{j\phi}E_{w,\mathrm{in}} -  \frac{k k^* \alpha e^{j (\theta_R+2\phi)}}{1-t  \alpha e^{j (\theta_R+\phi)}} E_{w,\mathrm{in}} = \frac{te^{j\phi}-t^2  \alpha e^{j (\theta_R+2\phi)} -k k^* \alpha e^{j (\theta_R+2\phi)} }{1-t  \alpha e^{j (\theta_R+\phi)}}E_{w,\mathrm{in}} =\\= \frac{te^{j\phi}- \alpha e^{j (\theta_R+2\phi)} }{1-t  \alpha e^{j (\theta_R+\phi)}} E_{w,\mathrm{in}}
\end{multline}

The field transmission coefficient of the whole device is thus given by:
\begin{equation}
\tau =e^{j\phi}\frac{t- \alpha e^{j (\theta_R+\phi)} }{1-t  \alpha e^{j (\theta_R+\phi)}}
\end{equation}
The power transmission coefficient on the other hand is the squared modulus of the former, namely:
\begin{multline}
\mathcal T = \left|\frac{t- \alpha e^{j (\theta_R+\phi)} }{1-t  \alpha e^{j (\theta_R+\phi)}}\right|^2 = \frac{(t- \alpha e^{-j (\theta_R+\phi)})(t- \alpha e^{j (\theta_R+\phi)})}{(1-t \alpha e^{-j (\theta_R+\phi)})(1-t  \alpha e^{j (\theta_R+\phi)})} =\\= \frac{ t^2+\alpha^2-t\alpha e^{j(\theta_R+\phi)}-t \alpha e^{-j(\theta_R+\phi)}}{1+\alpha^2 t^2-t \alpha  e^{-j(\theta_R+\phi)}-t \alpha e^{j (\theta_R+\phi)}} =  \frac{t^2+\alpha^2- 2 \,\Re \lbrace t \alpha e^{-j(\theta_R+\phi)}\rbrace}{1+\alpha^2 t^2- 2 \,\Re \lbrace t \alpha e^{-j(\theta_R+\phi)}\rbrace} 
\end{multline}
We observe that 
\begin{equation}
\Re \lbrace t \alpha e^{-j(\theta_R+\phi)}\rbrace  = t \alpha \cos(\theta_R+\phi)
\end{equation}
therefore we can finally write:
\begin{equation}
\mathcal T =  \frac{t^2+\alpha^2- 2  \alpha t \cos(\theta_R+\phi)}{1+\alpha^2 t^2- 2  \alpha t \cos(\theta_R+\phi)}  \label{eq:TresSI}
\end{equation}

\subsection{Coupling matrix for the MZI}

We now consider in more detail the form taken by the matrix $U$ in the case of a MZI, consisting in two-mode couplers with arbitrary phase shifts $\phi_1$ and $\phi_2$ in the arms in between. We assume the couplers to be equal, but not necessarily balanced, each namely described by a matrix:
\begin{equation}
U_{DC}=\begin{bmatrix}
\eta & j\sqrt{1-\eta^2} \\ j\sqrt{1-\eta^2}  & \eta
\end{bmatrix}
\end{equation}
with $\eta$ being a real parameter such that $0 \leq \eta \leq 1$. In fact, $\eta^2$ indicates the power transmission coefficient of the coupler on the same waveguide. The case of balanced couplers corresponds to $\eta=\sqrt{1-\eta^2}=\frac{\sqrt{2}}{2}$.

The matrix $U$ describing the full interferometer is written as:
\begin{align}
U
&= \begin{bmatrix} 0 & 1\\ 1 & 0\end{bmatrix}  \begin{bmatrix}
\eta & j\sqrt{1-\eta^2} \\ j\sqrt{1-\eta^2}  & \eta
\end{bmatrix} \begin{bmatrix} e^{j \phi_1} & 0 \\ 0 & e^{j \phi_2}\end{bmatrix} \begin{bmatrix}
\eta & j\sqrt{1-\eta^2} \\ j\sqrt{1-\eta^2}  & \eta
\end{bmatrix} = \notag \\
&= \begin{bmatrix} j \eta \sqrt{1-\eta^2} (e^{j \phi_1}+e^{j\phi_2}) & \eta^2 (e^{j \phi_1}+e^{j\phi_2}) - e^{j \phi_1} \\ \eta^2 (e^{j \phi_1}+e^{j\phi_2}) - e^{j \phi_2} & j \eta \sqrt{1-\eta^2} (e^{j \phi_1}+e^{j\phi_2})\end{bmatrix} 
 =\notag \\
&= e^{j\phi_{cm}} \begin{bmatrix} 2 \eta \sqrt{1-\eta^2} \cos\left(\frac{\Delta\phi}{2}\right) & -\sin\left(\frac{\Delta\phi}{2}\right)-j(2\eta^2-1)\cos\left(\frac{\Delta\phi}{2}\right) \\ \sin\left(\frac{\Delta\phi}{2}\right)-j(2\eta^2-1)\cos\left(\frac{\Delta\phi}{2}\right) & 2 \eta \sqrt{1-\eta^2} \cos\left(\frac{\Delta\phi}{2}\right)\end{bmatrix}  \label{eq:matrMZI}
\end{align}
where:
\begin{align*}
\Delta\phi &= \phi_1 - \phi_2 & \phi_{cm}&=\frac{1}{2}\left(\phi_1+\phi_2+\pi\right),
\end{align*}
Note that the matrix $\begin{bmatrix}
0 & 1 \\ 1 & 0 \end{bmatrix}$ has been added in the multiplication to keep  labeling of input and output fields consistent with Figure~\ref{fig:Fig_layout_description}. Comparing the matrix in \eqref{eq:matrRef} with the one \eqref{eq:matrMZI} we note that they have indeed the same form if we take:
\begin{align}
t&=2\eta\sqrt{1-\eta^2} \cos \left(\frac{\Delta\phi}{2}\right), & \phi&=\phi_{cm}.
\end{align}

In the case of ideal balanced couplers $\eta=\sqrt 2 /2$ we can write the cross-coupling coefficient of the MZI as:
\begin{equation}
u_{12} = u_{21} = t e^{j \phi_{cm}} = \cos \left(\frac{\Delta\phi}{2}\right) e^{j \phi_{cm}} \label{eq:balMZI}
\end{equation}
which is indeed Eq.~(1) of the Main Text.

We also observe that by writing $\phi=\phi_{cm}$ in \eqref{eq:TresSI}
we obtain precisely Eq.~(4) of the Main Text, which holds both in case of MZI with balanced couplers and in the more general case of unbalanced ones (considering the correct value for $t$).

\section{Fit of the transmission minima and maxima}
\label{secApp:Fit}

\begin{figure}[tbp]
    \centering
    \begin{tabular}{c}
    a)\includegraphics[width=.6\linewidth]{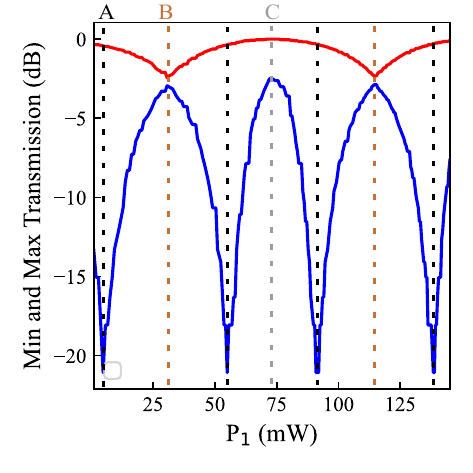}\\
    b)\includegraphics[width=0.8\linewidth]{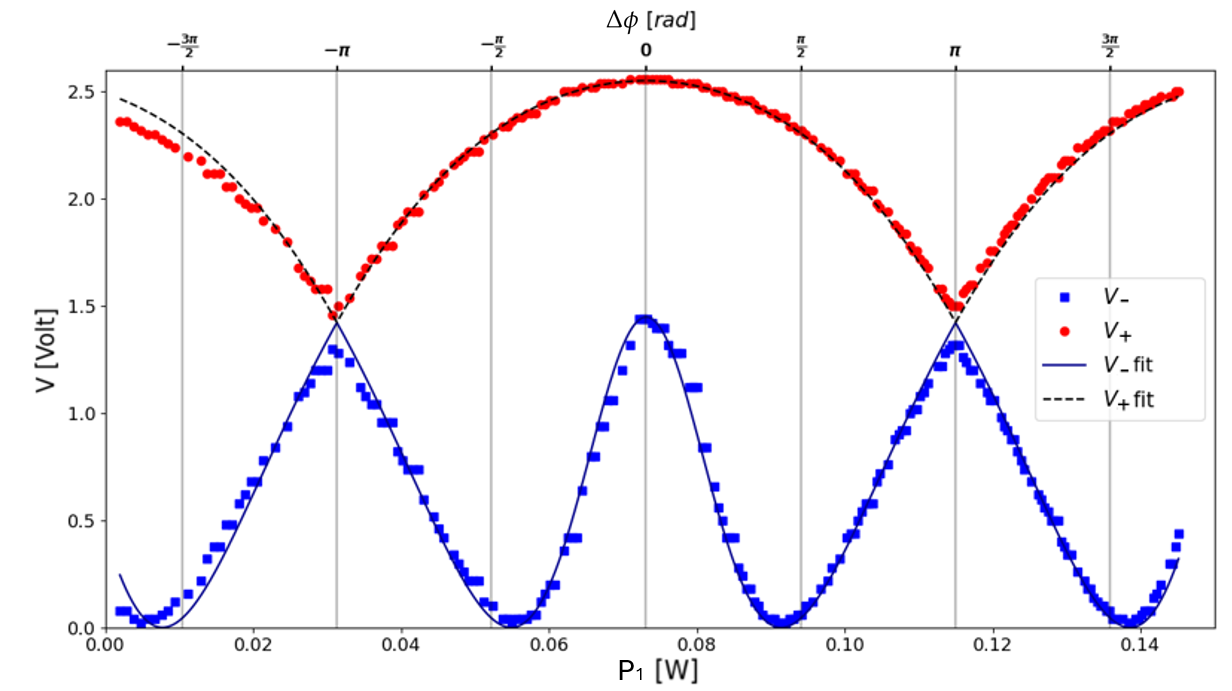}
    \end{tabular}
    \caption{\label{fig:Transmission}Maximum (red dot) and minimum (blue dot) transmission values extracted from the spectra as the voltage applied to $P_1$ is varied, thereby tuning the coupling coefficient $t$. Panel (a) plots data in logarithmic scale, which allows for a direct visualization of the extinction ratio and easy identification of notable phase shifts. In particular, the dotted lines labeled A (black), B (brown), C (grey) correspond respectively to the phases $\Delta \phi = \pm 2\arccos(\alpha)+2k\pi$,  $\Delta\phi=\pi+2k\pi$ and $\Delta\phi=0+2k\pi$, respectively. Panel (b) plots the same data in linear scale, together with best-fit curves for the maxima (dashed line) and minima (solid line).The upper horizontal axis reports the corresponding phase difference $\Delta\phi$, calculated from the fit.}
\end{figure}

Here we describe the fitting procedure used for the analysis of the experimental data in Fig. \ref{fig:Transmission} (corresponding to Fig. 2c in the Main Text). 

Firstly we recall that the resonator transmission, assuming two identical directional couplers with the same splitting ratio, is given by:
\begin{equation}
T = \frac{\alpha^2-4\eta \sqrt{1-\eta^2} \alpha \cos(\theta_R+\phi_{cm})\cos{\frac{\Delta\phi}{2}}+4\eta^2(1-\eta^2)\cos^2\frac{\Delta\phi}{2}}{1-4\eta \sqrt{1-\eta^2} \alpha \cos(\theta_R+\phi_{cm})\cos{\frac{\Delta\phi}{2}}+4\eta^2(1-\eta^2)\alpha^2\cos^2\frac{\Delta\phi}{2}},
\end{equation}
where $\eta$ is the amplitude self-coupling coefficient of the couplers.
For a fixed phase difference $\Delta\phi$, the maximum transmission ($T_{+}$) occurs at $\theta_R+\phi_{cm}=(2n+1)\pi$, for which the light entering the MZI interferes constructively at the output of the circuit. On the contrary the minimum ($T_{-}$) is observed at $\theta_R+\phi_{cm}=2n\pi$, for which destructive interference effectively traps light within the resonator. In a compact form, we can write:
\begin{equation}
T_{\pm} = \left(\frac{\alpha \pm 2\eta\sqrt{1-\eta^2}\cos\frac{\Delta\phi}{2}}{1 \pm 2\eta\sqrt{1-\eta^2}\alpha\cos\frac{\Delta\phi}{2}}\right)^2,
\label{T+T-}
\end{equation}
with the phase difference between the two MZI arms given by $\Delta\phi = a_1 P_1 + \Delta\phi_0$, if $P_2 $ is kept equal to zero. Finally, the measured voltage is proportional to the transmission, yielding the following fitting formulas:
\begin{equation}
V_{\pm} = c\left(\frac{\alpha \pm 2\eta\sqrt{1-\eta^2}\cos\frac{a_1P_1+ \Delta\phi_0}{2}}{1 \pm 2\eta\sqrt{1-\eta^2}\alpha\cos\frac{a_1P_1+ \Delta\phi_0}{2}}\right)^2,
\end{equation}
where $c$, $\alpha$, $\eta$, $\Delta\phi_0$, and $a_1$ are the fitting parameters. The primary goal of this procedure is to determine $\eta$ (thereby extracting the splitting ratio of the directional couplers) while $a_1$ and $\Delta\phi_0$ can be readily estimated from Fig.~\ref{fig:Transmission}a (see Section 2). Using the maximum and minimum values of the red data points, we can also determine $\alpha$ according to Eq.~7 in the Main Text.

From Eq. \ref{T+T-} we notice that, if the directional couplers are perfectly balanced (i.e. 50:50), for $\Delta\phi=0$ light exits the circuit without entering the ring section. Under these conditions, the transmission would be independent of the resonator length, with equal maximum and minimum values. However, in Fig. \ref{fig:Transmission} the blue and red points do not overlap for $\Delta\phi=0$, indicating a deviation from the ideal 50:50 balance. The fitting procedure used to extract $\eta$, namely the splitting ratio of the couplers, was performed on the data in linear Fig.~\ref{fig:Transmission}a, since Eq.~(16) refers to the linear (voltage) domain. In particular, the best-fit yields a 64:36 intensity splitting ratio.

\pagebreak

\section{Estimation of the linear coefficient $a_1$ and $a_2$}

\label{secApp:estCoeff}

From Eq.~(2) of the Main Text, the differential phase of an MZI with one thermo-optic phase shifter (TOPS) per arm can be expressed as:
\begin{equation}
\Delta \phi = \frac{2\pi}{\lambda} a_1 P_{1} +  \frac{2\pi}{\lambda} a_2 P_{2} + \Delta \phi_{0},
\end{equation}
where $a_1$ and $a_2$ quantify the contribution of the dissipated power in TOPS 1 and 2, respectively, and $\Delta \phi_{0}$ accounts for fabrication-induced phase offsets.

In order to preserve the critical coupling condition while shifting the resonance dip, it is necessary to activate $P_{2}$ to displace the dip, while simultaneously adjusting $P_{1}$ according to
\begin{equation*}
P_{1} = P_{1,\text{cc}} - \frac{a_2}{a_1} P_{2},
\end{equation*}
where $P_{1,\text{cc}}$ denotes the value of $P_{1}$ at critical coupling. 
This requires a preliminary calibration of the MZI to accurately determine the coefficients $a_1$ and $a_2$.

The calibration is carried out by activating each TOPS individually and scanning the dissipated power, which effectively tunes the coupling coefficient. 
Recalling the transmission coefficient of the MZI in the resonator with balanced couplers \eqref{eq:balMZI},
three operating points can be identified. 
The conditions $|t|=1$ %$t=1$
and $t=0$ occur at $\Delta\phi=0+2k\pi$ and $\Delta\phi=\pi+2k\pi$ corresponding to the maximum and minimum of the off-resonance resonator transmission (B and C point of Fig.~\ref{fig:Transmission}a). 
A third reference point is provided by the critical coupling condition, where the resonator is on resonance and the transmission reaches its minimum, referred to A in Figure \ref{fig:Transmission}a. 
In this case, $|t|^{2} = \alpha^{2}$ and the corresponding MZI phase is $\Delta \phi = \pm 2\arccos(\alpha)+2k\pi$. 

From Fig.\ref{fig:Transmission}a, we extract the powers $P_{0}$, $P_{\pi}$, and $P_{cc}$ corresponding to these three phase values. 
Using these reference points, the dissipated power can be related to the differential phase through a linear regression.
The same procedure is then repeated for the second heater. 
Owing to the symmetric geometry of the two resistors located on opposite interferometer arms, one expects coefficients of comparable magnitude but opposite sign.

\section[Quality factor and propagation losses]{Quality factor and propagation losses}

\label{secApp:Qfactor}

In the limit of a sufficiently high finesse (typically $>3-4$), it is possible to relate the intrinsic quality factor of a ring resonator to the losses per unit length $\alpha_L$ associated with the propagation in the resonator waveguide. Such losses can be due to scattering, material absorption, or bending losses. We start by considering the definition of the intrinsinc quality factor as 
\begin{equation}\label{quality_factor}
    Q_i=\frac{\omega_0}{\gamma_i},
\end{equation}
where $\omega_0$ is the resonant frequency and $\gamma_i$ is the intrinsic radiation loss rate, which can be written as
\begin{equation}\label{dissipation_rate}
    \gamma_i=\frac{1-e^{-\alpha_L \mathcal{L}}}{\tau_\mathrm{rt}}\approx\frac{\alpha_L \mathcal{L}}{\tau_\mathrm{rt}}=\alpha_Lv_g,
\end{equation}
where $\mathcal{L}$ is the round-trip length and $\tau_\mathrm{rt}$ is the round-trip time, and where we used $\tau_\mathrm{rt}=\mathcal{L}/v_g$, being $v_g$ the group velocity.
By inserting \eqref{dissipation_rate} in \eqref{quality_factor}, we obtain
\begin{equation}\label{Q_i}
    Q_i=\frac{\omega_0}{\alpha_Lv_g}=\frac{2\pi n_g}{\lambda_0\alpha_L}.
\end{equation}
Finally, we notice that $Q_\mathrm{cc}=Q_i/2$.

\end{document}